# MOND and the Galaxies

F. Combes[a] and O. Tiret[b]

[a]*Observatoire de Paris, LERMA, 61 Av de l'Observatoire,*
*F-75014 Paris, France*
[b]*SISSA, via Beirut 4, I-34014 Trieste, Italy*

**Abstract.** We review galaxy formation and dynamics under the MOND hypothesis of modified gravity, and compare to similar galaxies in Newtonian dynamics with dark matter. The aim is to find peculiar predictions both to discriminate between various hypotheses, and to make the theory progress through different constraints, touching the interpolation function, or the fundamental acceleration scale. Galaxy instabilities, forming bars and bulges at longer term, evolve differently in the various theories, and help to bring constraints, together with the observations of bar frequency. Dynamical friction and the predicted merger rate could be a sensitive test of theories. The different scenarios of galaxy formation are compared within the various theories and observations.

## I- INTRODUCTION

Milgrom (1983) has proposed an empirical modification of the law of gravity, as an alternative to dark matter in galaxies. The basic idea of this MOdified Newtonian Dynamics (MOND) derives from the observation that the missing mass problem occurs only in the weak field regime, at low acceleration (see Figure 1). In addition, galaxies are following the baryonic Tully-Fisher relation (McGaugh et al 2000), which tells us that the maximum rotational velocity of a galaxy runs as the $1/4^{th}$ power of its mass, and does not depend on the radius, suggesting an asymptotic acceleration in $1/r$. The proposition is thus that at acceleration below $a_0 = 10^{-10}$ m/s$^2$, the gravitational attraction will tend to the formulation $a = (a_0 \, a_N)^{1/2}$, where $a_N$ is the Newtonian value. This effectively produces an acceleration in $1/r$, implying a flat rotation curve in the limiting regime, and automatically the Tully-Fisher relation. The transition between the Newtonian and MOND regime is controlled by an interpolation function $\mu(x)$, of $x = a/a_0$, which standard form is $\mu(x) = x/(1+x^2)^{1/2}$, that essentially tends to x in the MOND regime, when x is smaller than 1, and to unity in the Newtonian regime. This phenomenology has a large success explaining rotation curves and kinematics of galaxies, from irregular dwarf dominated by dark matter (and therefore in the MOND regime), to the giant and ellipticals, dominated by baryons (e.g. the review Sanders & Mc Gaugh 2002).

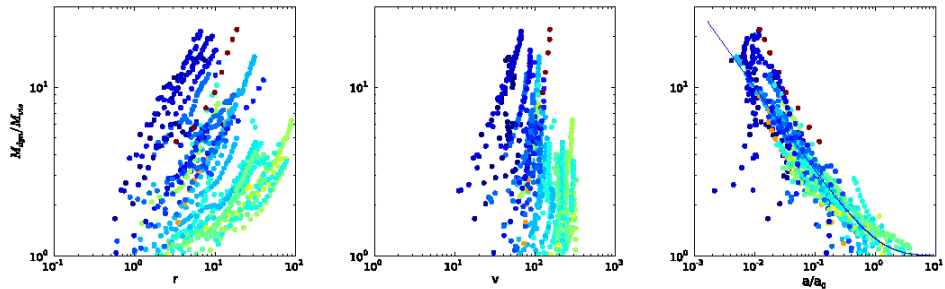

**FIGURE 1.** The missing mass problem. In the three panels, the dynamical mass to visible mass ratio is plotted versus radius (left), velocity (middle), and acceleration (right), for a series of nearby galaxies, with clean HI rotation curves. Only the acceleration reduces the scatter (cf Tiret & Combes 2009).

The motivation for MOND is not only an alternative to dark matter, which particles have not yet been found, but also to solve the problems that the standard CDM model encounters at galaxy scales. Numerical simulations in the standard model predict an over concentration of dark matter in galaxies, and cuspy density profiles, instead of the density cores derived from rotation curves (e.g. de Blok et al 2008, Swaters et al 2009). Also simulations have difficulties to form large galaxy disks, since the angular momentum of baryons is lost against massive dark haloes (e.g. Navarro & Steinmetz 2000), and the missing satellites problem is more acute as spatial resolution is increased (Diemand et al 2007). Possible solutions of these problems may reside in the detailed physics of the baryonic component, that is not presently reproduced realistically enough in simulations: feedback from star formation, supernovae and galaxy-scale winds, or from energy released by the central active nucleus, are currently investigated. In parallel, it is interesting to explore what are the constraints on the MOND phenomenology, due to galaxy dynamics, confronted to observations.

## II- CONSTRAINTS ON MOND FROM GALAXY DYNAMICS AND OBSERVATIONS

Reproducing the kinematics of galaxies is not sufficient to test a model. More constraints can be obtained from the stability, evolution and formation of galaxies. In MOND, disks are totally self-gravitating, and not embedded in dark spheroidal halos, which usually do not rotate and are not subject to instabilities. Disks could then be more unstable, however, gravity tends to be proportional to the square root of mass only, which might have a limiting effect. Only simulations can provide an insight on the dominating effect.

### II-1. Determination of the interpolation function

In a large number of nearby galaxies, the best interpolating function $\mu(x)$ is the standard form $\mu(x) = x/(1+x^2)^{1/2}$, (e.g. Sanders & Verheijen 1998, Milgrom & Sanders 2007), however, the simple one, $\mu(x) = x/(1+x)$ has also been used, being consistent with the TeVeS (Tensor, Vector, Scalar) relativistic extension of MOND (Bekenstein 2004). The simple function gives also good fits, but with a lower mass-to-light ratios for the stellar component (Sanders & Noordermeer 2007). In the Milky Way, however, it appears that the standard form has some difficulties to fit all the data, and the simple form is better, although another third form would be optimum (Famaey & Binney 2005). It is to be kept in mind, however, that being inside the Galaxy, makes distances and actual rotating velocities highly uncertain. One fact is certain, the Milky Way is dominated by baryons all across the visible disk, contrary to the predictions of the CDM model.

### II-2 Escape Velocity

The potential around a galaxy of mass M in the pure MONDian regime is logarithmic $\Phi(r) = (GMa_0)^{1/2} \ln r$, and there should not be any escape possible. However, a galaxy is never totally isolated, and the companions provide an external field, which modifies totally the shape of the potential. This is called the External Field Effect (EFE). When solving the Poisson equation for MOND, div $(\mu(x) \mathbf{g}) = -4\pi G\rho$ (X, Y, Z), with $x=g/a_0$, and assuming an external field $g_e$ in the X direction, for instance, it is possible to define an equivalent internal potential $\Phi_{int}$, which at large radii, is equivalent to a dilatation $\Delta$ in the direction transverse to the external field):
$$\Phi_{int} = -GM/\mu_m (X^2 + (1+\Delta)(Y^2+Z^2))^{-1/2}$$
Here $\mu_m$ is the interpolation function corresponding to the acceleration of the centre of mass. This allows to compute the escape velocity. In the case where $g \ll g_e \ll a_0$, the shape of the potential has retrieved a Keplerian dependence, with the renormalization of the gravitation constant G into $Ga_0/g_e$.

The application has been done to the Milky Way, where recent observations have constrained the escape speed between 498 and 608 km/s (Smith et al 2007). The most important external field effect comes from Andromeda, with $g_e = a_0/100$. Simulations have been done with the Besançon mass model of our Galaxy, and provide a good agreement, as shown in Figure 2 (Wu et al 2007).

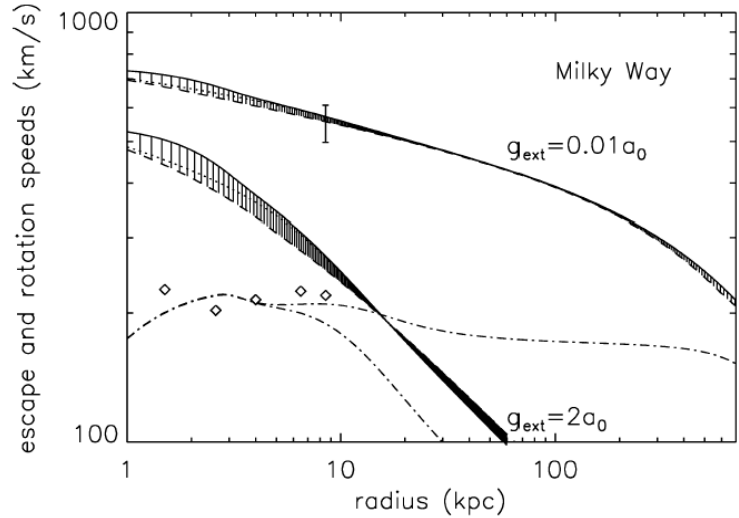

**FIGURE 2.** The influence of a weak ($0.01a_0$) and strong ($2a_0$) external field is computed on the Milky Way Galaxy. The full lines and hashed zones are the escape speed, for various field directions, compared with the error-bar measured by RAVE (Smith et al 2007). The dash-dot lines are the rotation velocity, compared by the symbols from Caldwell & Ostriker (1981). The weak field provides a good fit to the observations (from Wu et al 2007).

In addition, the EFE has some unexpected consequences, such as disk precession. If the acceleration ge can be considered as constant in amplitude and direction during the rotational period of the central disk, then it implies a gravitational torque and precession around this direction. This is completely absent in the Newton case, where only the tidal effect, and therefore the variation of ge is felt over the disk. This is due to the non-linearity of MOND, which violates the strong equivalence principle. Simulations show that the differential precession over the disk provokes warps (Tiret & Combes, in prep), and this could provide a new interpretation for isolated warped galaxies (see also Brada & Milgrom 2000).

## II-3 Stability of galaxy disks: spirals and bars

Spiral waves and bars are the motor of evolution. While CDM dark haloes at first stabilise galaxy disks, they can also provide amplification of bars, since they accept angular momentum: bars are negative momentum waves, they grow through outward transfer of angular momentum. It is therefore interesting to test the stability and bar frequency in both hypotheses.

N-body simulations have resolved the Poisson equation for spiral disks, through a multi-grid algorithm (Tiret & Combes 2007). Initial conditions in phase-space for the visible galaxies were exactly the same, for the Newtonian and MOND cases. Once the baryonic disk had been selected, and provided a realistic rotation curve in MOND, then the dark matter radial distribution was computed, in the Newtonian case, in order to provide exactly the same kinematics, i.e. the same phase-space distribution for the baryons. Figure 3 shows the different results found, in the case of pure stellar systems. With CDM, the bar appears later, and can reform after the first bar is weakened due to the peanut formation (vertical resonance). With MOND, the bar appears quickly through angular momentum transfer to the outer disk, and stays longer, but it does not re-appear, by lack of material to accept more angular momentum. Besides, the pattern speed of the bar behaves very differently: in CDM, the bar is slowed down by dynamical friction against the halo, while the pattern speed stays constant in MOND. The latter is more consistent with observations, that reveal fast bars, through the position of resonances.

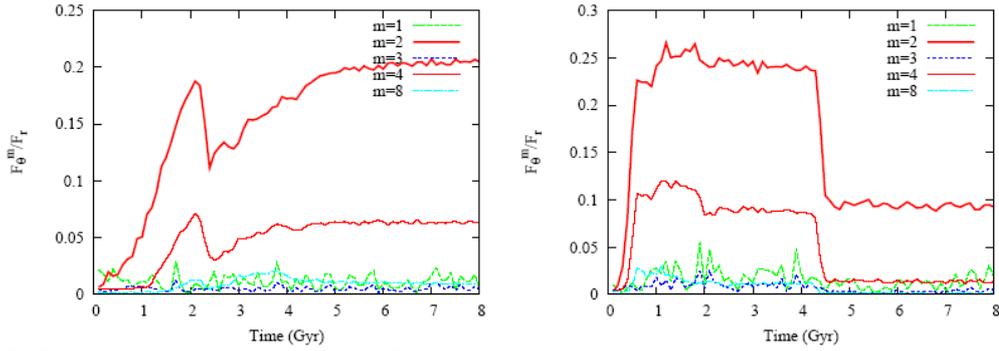

**FIGURE 3.** Strength of the bar, developed in an Sa galaxy stellar simulation, measured by its Fourier harmonics m=2,3,4 and 8, for the CDM-Newton model (left) and MOND (right). The bar settles earlier in MOND, and stays longer, but after its drop at 4.5 Gyr, it does not develop again, as in the CDM (cf Tiret & Combes 2007).

The slowing down of the bar in the CDM case is accompanied by several peanut formations. The vertical resonance occurs at larger and larger radii, while the bar pattern speed decreases. In MOND, the peanut does not move in radius.

The statistics of the bar strength, taking into account all galaxy models across the Hubble sequence, have been compared with observations. When the gas component is included, a large variety of situations are encountered (gas inflows can destroy bars, external gas accretion can reform them, Bournaud & Combes 2002, Combes 2008), and both models are compatible to the data. Gas radial flows due to the bar gravity torques are responsible to the peculiar morphology of spiral disks: gas piles up at resonances, and form new stars in rings or pseudo-rings. Simulations show that these dynamical phenomena in MOND compatible with observations (cf Figure 4).

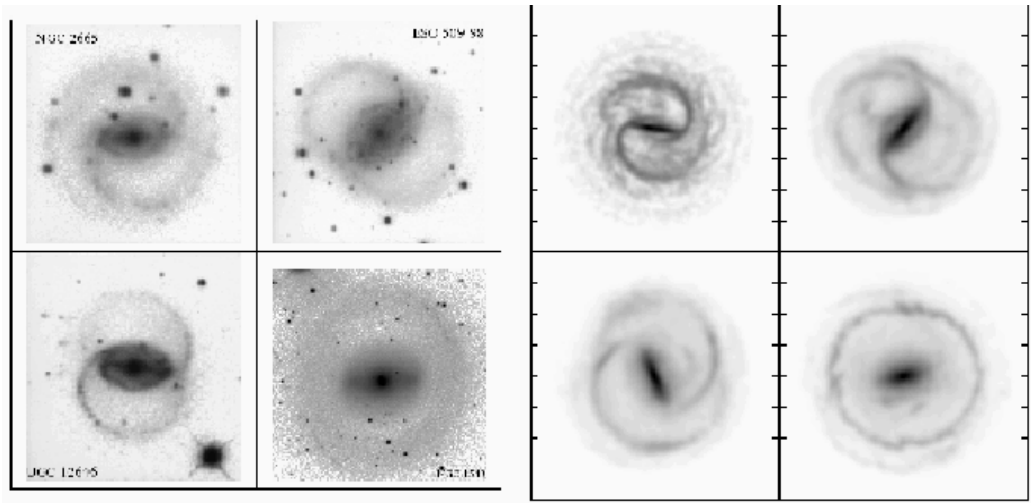

**FIGURE 4.** Formation of resonant rings through the bar in MOND simulations with gas and star formation. The 4 images at left are particular galaxies to confront with the 4 simulated images at right. MOND can reproduce realistically the morphology of galaxy disks (cf Tiret & Combes 2008a).

## II-4 Interactions of galaxies, dynamical friction

Galaxy interactions and mergers are a fundamental element of the hierarchical scenario of galaxy formation. At large distance, the interactions are likely to occur in the MOND regime, and extended dark haloes are essential to absorb the orbital angular momentum for two galaxies to merge. Ciotti & Binney (2004) notice that force fluctuations are larger in the MOND regime, and from a perturbation estimation of the dynamical friction time, conclude that dynamical friction should be much more efficient in MOND, for instance for stellar clusters spiralling inward or bar slowing down. Mergers should also occur more quickly. However, numerical simulations reveal the opposite effect. Mergers time are much longer in MOND. This has been obtained by merging dissipationless spherical systems with a polar code (Nipoti et al 2007), and also merging spiral galaxies (disks with gas) like the Antennae system, the prototype of major mergers (Tiret & Combes 2008b).

Figure 5 shows that MOND is able to reproduce the Antennae as well as the CDM model. Simulations are done here with an adaptive mesh, easy to implement with multi-grid methods. Dynamical friction is very slow in MOND, since galaxies are not embedded in massive spheres of particles, able to accept the orbital angular momentum. A short merging time-scale, as short as the CDM, is possible, only for nearly radial orbits.

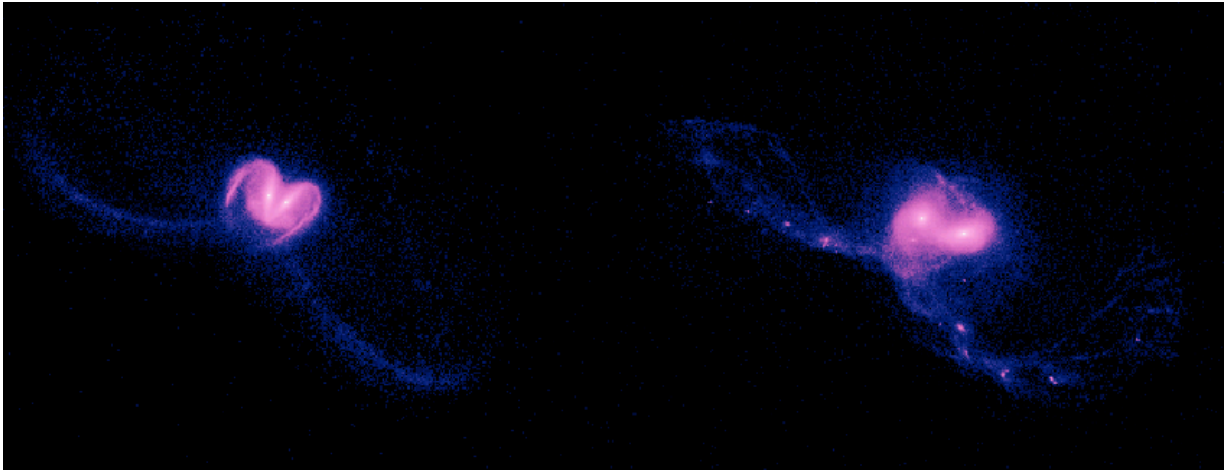

**FIGURE 5.** Simulations of the Antennae galaxies with CDM-Newton at left, and MOND at right. The morphological details have not been fitted, but the main features and tidal tails are reproduced (cf Tiret & Combes 2008b).

Why then dynamical friction has been predicted analytically to be stronger with MOND than in the ENS (Equivalent Newtonian System) with dark matter (Ciotti & Binney 2004)? Both galaxy interactions or bar slowing down have shown the opposite. This must be due to the approximations done of weak perturbations. Force fluctuations are indeed stronger in MOND, but when a massive body (either a companion, or a galactic bar) has to be slowed down, the particles in the outer disk of galaxies are not sufficient to absorb large amounts of momentum. The reservoir of particles, assumed infinite in the case of a weak perturbation, is insufficient. In the CDM model, the reservoir of dark matter particles is found in the assumed dominating CDM component (cf Nipoti et al 2008).

*Merger induced starbursts degeneracy*

In the standard model, dynamical friction on DM particles is very efficient, and major mergers between two spiral galaxies in a parabolic orbit are expected to occur in one or two passages. An intense starburst is associated to the final phases, and the number of starbursts is thought to count the number of mergers. In the MOND model, many passages in binary galaxies will be required before the final merging. However, a starburst may be triggered at each passage (e.g. Di Matteo et al 2007). The number of starburst, as a function of redshift, could then be similar, and cannot discriminate the two models. The number of apparently merging galaxy in the sky is also degenerate. There could be a limited number of long-lived mergers, or many more short-live mergers.

## II-5 Formation of Tidal Dwarfs

Major mergers between spiral galaxies are frequently observed with young dwarf galaxies at the extremity of their tidal tails, called Tidal Dwarf Galaxies (TDG). In the CDM model, they are difficult to form, and require very extended DM distribution (Bournaud et al 2003). In MOND, the exchange of angular momentum occurs within the disks, which sizes are inflated. It is then much easier with MOND to form TDG. The formation of new galaxies out of tidal debris is the occasion to test the missing mass problem. They are not expected to drag enough CDM, and should present no dark matter problem. The rotation curve of three TDG in the NGC 5291 ring system however, reveals the presence of dark matter (Bournaud et al 2007). A solution with the standard model is to resort to dark baryons in the form of cold molecular gas (e.g. Pfenniger et al 1994). These TDG are well explained in the MOND phenomenology (Gentile et al 2007, Milgrom 2007). However, three TDG are not yet enough statistically to consolidate the result, inclinations on the plane of the sky are uncertain, and selecting a higher inclination eliminates the missing mass. Many other TDG should be observed to randomize the uncertainties.

## II-6 MOND and the dark baryons

Since stars and visible gas in galaxies represent only 6% of all baryons, a large fraction of dark baryons must exist outside the galaxies in filaments, and certainly a small fraction of them in galaxies under a cold gas form. Is MOND compatible with the existence of dark gas in galaxies? What fraction of dark gas provides the best fit to the rotation curves? A recent modelling of about 50 rotation curves, introducing dark gas and allowing a0 to vary accordingly, has shown that the optimum gas fraction was c=M(dark)/MHI= 2-3 (Tiret & Combes 2009). Figure 6 shows the quality of the fits for different values of this ratio, for the dwarf Irr NGC 1560. The fact that there is some sort of degeneracy here, comes from the long known observation that dark matter and HI surface densities are proportional in spiral galaxies (e.g. Hoekstra et al 2001).

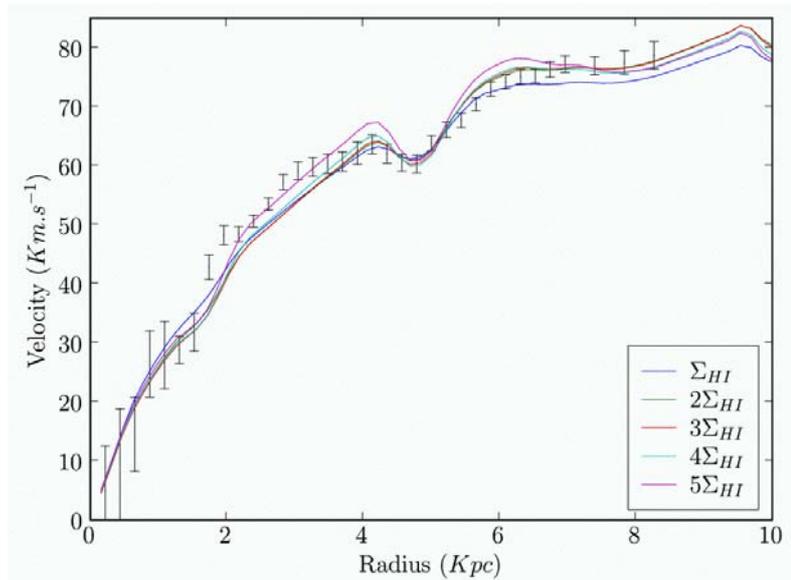

**FIGURE 6.** Several models for the rotation curve of the dwarf Irregular galaxy NGC 1560. The various curves correspond to the hypothesis that the total gas is either only the HI observed ($\Sigma_{HI}$), or a multiple of it, taking into account the dark molecular gas. In each case, the value of a0 is lowered accordingly (cf Tiret & Combes 2009).

## II-7  MOND and Ellipticals

Elliptical galaxies are hot systems, with little or no cold gas, and determining the amount of missing mass is much more difficult than for spirals. Several tracers have been used, such as X-ray gas, globular clusters, planetary nebulae, which revealed a possible dearth of dark matter (Romanowsky et al 2003). However, the velocity dispersion could be decreasing only because of radial orbits in the outer parts. In particular, a merger between two spirals is expected to form an elliptical, with radial orbits in the merging direction (Dekel et al 2005). In general, the velocity anisotropy $\beta = 1 - \sigma^2_\theta/\sigma^2_r$ introduces a strong degeneracy, $\beta$ can take large negative values (circular orbits), =0 (isotropic) to 1 (radial orbits). Figure 7 shows some models in the frame of MOND for a well studied elliptical NGC 3379.

## II-8 Missing mass at large scale from satellites in SDSS

There is not a lot of tracers of the dark matter at larger scale. Recently Klypin & Prada (2009) have computed statistically the velocity dispersion of satellites around early-type galaxies, in the Sloan Survey (SDSS). There is only one or zero satellite for each galaxy in average, but thousands of galaxies. Again, the anisotropy of velocities introduce degeneracy in the models, CDM and MOND can both reproduce the data, with different values of β (Figure 7, see Angus et al 2008), however the CDM standard model with cusps is not a good fit.

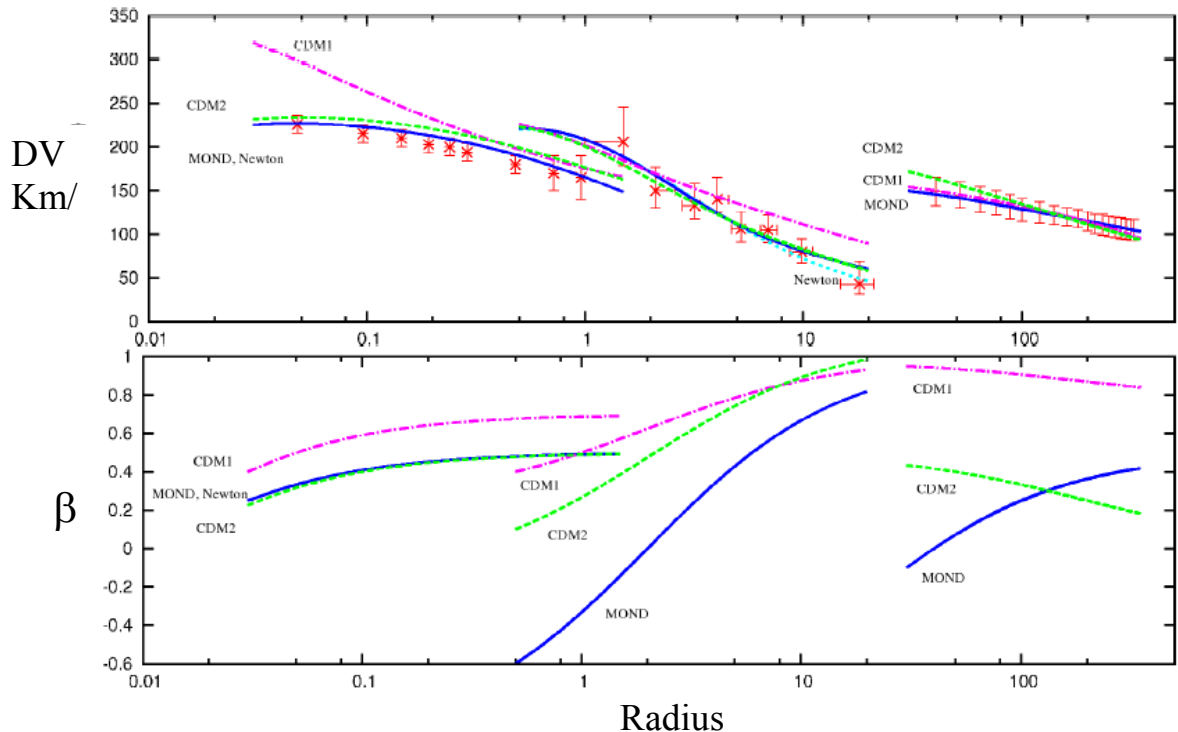

**FIGURE 7.** Velocity dispersion around the elliptical galaxy NGC 3379 (data from Shapiro et al 2006 in the centre, and Douglas et al 2007 at 10-30kpc), and on the same graph, large-scale mass distribution statistically derived from satellite data around ellipticals (data from Klypin & Prada 2009). The curves correspond to the various models: Newton model without DM, two different radial distributions of CDM (CDM1 with cusps, and CDM2 with cores), and MOND. Fits can be obtained in varying the velocity anisotropy β, shown in the bottom (cf Tiret et al 2007).

## II-9 Large scale structure

Cosmological simulations with the MOND gravity have been explored, to test its ability to reproduce the large-scale structure formation, which is one of the best successes of the CDM theory. Previous works had made the strong approximations that the Newton and MOND acceleration fields were parallel (Nusser 2002, Knebe & Gibson 2004). They started from a cosmological Newtonian with CDM universe, assumed that the critical a0 was constant, and found that MOND produced as much clustering. Releasing the parallel approximation, Llinares et al (2009) have carried out new simulations, assuming Newtonian initial state, and a critical acceleration a0 varying with time, proportional to the scale factor of the universe. Their simulations yield comparable results between MOND and the standard model for the large-scale structure, with even more clustering than with the parallel approximation. The cosmological models tested are open with only baryons (without dark energy), and only gravity forces are considered (no hydrodynamics). There is still much freedom and many models to explore, within the MOND frame.
.

# III- CONCLUSION

Recent models and simulations demonstrate that MOND phenomenology is able to reproduce observations of galaxy dynamics, and solve the problems of CDM at galaxy scales. The bar frequency is compatible with what is observed, and the number of mergers/starbursts is degenerate. Dynamical friction is much smaller than in the CDM model, and this could explain the frequency of compact groups.

However, MOND encounters remaining missing mass problems at cluster scale, which are not yet resolved, although neutrinos or dark baryons have been invoked.

The exploration of bar formation and galaxy interactions in MOND has suggested that more observational tests could be carried out to better constrain the various MOND models. In particular the dynamics of tidal debris should be studied statistically.